# Accelerating Battery Material Optimization through iterative Machine Learning

Seon-Hwa Lee[1], Insoo Ye[1], Changhwan Lee[1], Jieun Kim[1], Geunho Choi[1], Sang-Cheol Nam[1], Inchul Park[1, *]

[1] POSCO N.EX.T Hub, Incheon, Korea

[*] Corresponding author at: inchul@posco-inc.com

## Abstract

The performance of battery materials is determined by their composition and the processing conditions employed during commercial-scale fabrication, where raw materials undergo complex processing steps with various additives to yield final products. As the complexity of these parameters expands with the development of industry, conventional one-factor-at-a-time (OFAT) experiment becomes old fashioned. While domain expertise aids in parameter optimization, this traditional approach becomes increasingly vulnerable to cognitive limitations and anthropogenic biases as the complexity of factors grows. Herein, we introduce an iterative machine learning (ML) framework that integrates active learning to guide targeted experimentation and facilitate incremental model refinement. This method systematically leverages comprehensive experimental observations, including both successful and unsuccessful results, effectively mitigating human-induced biases and alleviating data scarcity. Consequently, it significantly accelerates exploration within the high-dimensional design space. Our results demonstrate that active-learning-driven experimentation markedly reduces the total number of experimental cycles necessary, underscoring the transformative potential of ML-based strategies in expediting battery material optimization.

# Introduction

The accelerated development and widespread commercialization of battery technologies has significantly increased the complexity and scope of materials development [1-5]. As performance requirements become increasingly challenging and economic pressures intensify, researchers and engineers are now faced with an extensive and intricate parameter space involving material composition, processing variables, and external economic and regulatory considerations. [6-9] Furthermore, the entire manufacturing chain—from raw material mining and refining, through material manufacturing, to final cell assembly—adds another layer of complexity, as each stage introduces distinct parameters and interdependencies that can profoundly influence performance and cost. [10-12] (**Figure 1a**) Traditional one-factor-at-a-time (OFAT) experimental methods, despite their simplicity and intuitive appeal, are increasingly insufficient for capturing complex interactions among numerous parameters. [13,14]

For instance, the fabrication of nickel-cobalt-manganese (NCM) cathode materials involves multiple critical parameters, including precursor composition (e.g., elemental ratio of Ni, Co, Mn), morphology and its distribution, quantitative ratios of Li precursor to metal precursor, individual particle sizes of both precursors, doping or coating methods and the type and amount of additives employed, crystal structure and particle size in solid-state synthesis, and synthesis atmospheric conditions (schematically illustrated in **Figure 1b**). Each of these parameters not only substantially impacts electrochemical performance but also exhibits complex interdependencies, making it challenging to isolate and optimize their individual contributions. Conventional heuristic optimization methods, which rely heavily on expert domain knowledge and intuitive transfer learning from existing materials systems, often struggle to systematically explore and accurately pinpoint optimal parameter combinations. [14] Consequently, given the rising complexity of performance criteria, cost sensitivity, and regulatory constraints, there is a clear and pressing need for more sophisticated, systematic, and data-driven approaches to address these evolving demands in battery material development.

Integrating machine learning (ML) with systematic experimental design has emerged as a promising solution capable of overcoming the limitations of traditional approaches. [15-21] ML methodologies are uniquely suited to handle extensive, high-dimensional parameter spaces, effectively uncovering complex interactions and hidden patterns that conventional Edisonian methods frequently overlook. Nevertheless, the successful application of ML in battery material optimization is often impeded by data scarcity, incomplete metadata, and systematic biases present in available datasets. [22,23] Although high-throughput experimentation (HTE) combined with materials informatics has partially mitigated these challenges by significantly increasing data volumes, these methods currently struggle to accurately replicate or represent the complexities of actual industrial-scale processes. [24] Moreover, while HTE methods improve data availability, they do not inherently eliminate anthropogenic biases or address the chronic underreporting of negative outcomes. [25] Indeed, recent research emphasizes the critical importance of systematically incorporating "failed" or suboptimal experiments into ML training datasets, thus fully capturing the design space and enhancing model robustness. [26] Such strategies facilitate a shift from traditional Edisonian trial-and-error methods toward rigorous, data-centric frameworks.

In this context, active learning—which iteratively refines ML models by strategically selecting new experiments—offers a particularly promising approach. [27-30] By deliberately including experiments that might otherwise be dismissed by conventional heuristic methods, active learning systematically reduces anthropogenic biases and provides deeper insights into the parameter interactions important for performance. Prior research in inorganic and catalytic materials has illustrated how even a modest number of well-chosen active learning-driven experiments can dramatically improve predictive accuracy and model interpretability, guiding researchers toward optimized material formulations that conventional heuristic methods could easily overlook. [31] Motivated by successful demonstrations across diverse materials systems, we adapted an active learning framework specifically to address data limitations within real-world battery material processes.

In this study, we propose and validate an iterative, iterative ML-driven active learning framework aimed at optimizing NCM-based cathode materials under realistic processing conditions. A systematically curated dataset was developed through targeted NCM synthesis experiments explicitly designed to include traditionally underreported suboptimal outcomes. Through iterative training and strategic experiment selection guided by our ML framework, we successfully enhanced the robustness and accuracy of predictive models. Our results demonstrate the effectiveness of active learning in navigating the complexities of real-world battery manufacturing processes, revealing previously hidden parameter interactions and significantly accelerating the discovery and optimization of novel battery materials.

# Results and discussion

**Anthropogenic Bias in Heuristic-Driven Experimental Records: NCM cathode case**

To illustrate the limitations of conventional OFAT approaches and leverage ML's benefits, we compiled a comprehensive dataset from years of heuristic-driven NCM cathode synthesis research. This dataset, while large, was not originally intended for ML, thus embedding potential anthropogenic biases introduced by repeated human decision-making. [32] It encompasses key design parameters (*e.g.*, elemental composition, coating materials, and pivotal process parameters such as synthesis and coating temperatures. see **Figure 1b** for a detailed overview).

Notably, consistent experimental conditions were meticulously maintained wherever possible, aiming to minimize the potential influence of confounding factors beyond the explicitly varied design parameters. Electrochemical performance of battery materials is sensitive not only to lattice-level factors such as elemental composition and dopant chemistry but also significantly to micro-scale uniformity. [33-35] For instance, variations in mixing homogeneity, heat distribution uniformity in the furnace, and energy transfer driven by gas flow can profoundly impact the resulting material properties. [36,37] Therefore, preserving uniformity in these factors across experiments was crucial to clearly isolate and evaluate the effects solely attributable to intentionally modified design variables.

Despite carefully controlling experimental conditions to ensure consistency and facilitate predictive modeling, preliminary statistical analysis of the collected data revealed critical issues limiting its direct applicability for conventional machine learning approaches. First, the dataset exhibited a clear over-representation of successful experiments, whereas suboptimal or unsuccessful experiments tended to be partially incomplete. Specifically, experiments yielding poor initial electrochemical performance were often prematurely discontinued, resulting in missing data for subsequent characterization steps such as structural analyses or imaging. This partial incompleteness effectively led to systematic underreporting of suboptimal outcomes, aligning with concerns previously raised in the literature, [32,25] which emphasize that insufficient reporting of negative or incomplete data can significantly impair ML models by depriving them of crucial boundary conditions.

Second, a pronounced anthropogenic bias was identified within the dataset distributions. As illustrated in **Figure 2**, several critical parameters exhibit markedly skewed distributions. For example, while the synthesis temperature follows a relatively Gaussian distribution due to systematic variations across trials, the coating temperature data is tightly clustered around a single predominant value. Interviews conducted with researchers revealed that this clustering primarily resulted from heuristic decision-making, wherein researchers preferentially set the coating temperature close to the known melting point of the selected coating material. [38] However, actual melting behavior in practice can deviate significantly due to complex multi-phase interactions and compositional variations. Consequently, restricting experiments predominantly to this narrow temperature region could inadvertently neglect other potentially optimal or superior conditions. Such anthropogenic biases thus fundamentally limit the breadth and exploratory power of the collected dataset. [32]

An additional aspect of anthropogenic bias became evident in the persistent fixation of certain parameters that were initially deemed optimal at an early stage of the research. Several process variables exhibited extreme "one-value" distributions, indicating a strong preference by researchers to maintain previously successful parameter values rather than systematically exploring alternative conditions in subsequent experiments. This heuristic-driven fixation significantly constrained the dataset by reducing the exploration of potentially optimal conditions, thus limiting comprehensive parameter-space coverage. Consequently, the resulting data imbalance and limited variability pose substantial challenges to constructing robust and generalizable machine learning models capable of accurately predicting material performance beyond narrowly defined experimental conditions.

**Baseline ML and the Need for Active Learning**

To establish a baseline for predictive modeling, we initially tested several conventional ML models—including Decision Tree (DT), Random Forest (RF), Gradient Boosting Machine (GBM), and Neural Network (NN)—to predict the initial discharge capacity solely from our existing dataset ("zero-shot"). Table A summarizes their baseline predictive performances in terms of RMSE, $R^2$, and MSE. Among the tested models, GBM demonstrated the highest accuracy, a result consistent with previous findings indicating that ensemble-based methods generally exhibit superior robustness in handling noise and skewed distributions. [39-41]

Despite the GBM model's strong predictive accuracy within the narrowly defined chemical and process parameter space already captured by our existing dataset (**Figure 3a**), its predictive performance quickly plateaued when attempting to generalize beyond well-represented conditions. Specifically, the model tended merely to reproduce known successful parameter combinations without providing new insights or predicting outcomes beyond the experimentally well-explored regions. This limitation clearly highlights a fundamental shortcoming of applying conventional ML to datasets with inherent biases or incomplete parameter coverage: while these models excel at interpolating known data points, they are notably deficient in identifying novel, potentially optimal parameter spaces. Therefore, to overcome this limitation and systematically explore previously neglected or underrepresented conditions, we subsequently implemented an iterative active-learning framework designed specifically to enhance parameter-space exploration and predictive generalization.

To address the constraints of data scarcity and anthropogenic bias, we implemented an active learning loop guided explicitly by ML predictions. Given the dataset was predominantly composed of experiments at approximately 92% Ni content (634 samples), further improvements in this extensively explored region appeared limited. In contrast, higher Ni-content regions, particularly around 94% Ni (18 samples), remained substantially underrepresented, presenting a clear target for expansion through systematic exploration.

**Iterative Active Learning: Expanding Parameter Space and Improving Predictions**

Initially, we utilized our baseline GBM model—trained solely on the existing dataset—to identify parameter combinations predicted to yield the highest electrochemical performance, specifically focusing on the less-explored 94% Ni composition and previously untested conditions. Unlike traditional heuristic-based experimental selection, this ML-driven approach objectively prioritized experiments based purely on predictive outcomes derived from the current model. New experiments were subsequently performed under these ML-suggested conditions, and the resulting data—encompassing both high-performing and suboptimal outcomes—were incrementally incorporated into the model through retraining.

As illustrated in **Figure 3b**, iterative active learning significantly enhanced the predictive accuracy of our model and systematically revealed high-performance parameter regions that had previously remained underexplored. In the initial ("one-shot") iteration, the GBM model predicted a maximum discharge capacity of approximately 226 mAh/g for newly targeted conditions, and subsequent experimental validation across ten selected conditions yielded actual capacities ranging from 215 to 229 mAh/g. Although promising, this initial iteration still exhibited notable variance, with an average predictive error of approximately 2.5%.

Following retraining that incorporated both successful and suboptimal experimental results from the first iteration, the second iteration ("two-shot") demonstrated further refined predictions, forecasting an increased maximum discharge capacity of around 229 mAh/g. (see **Figure 4**) Subsequent validation experiments conducted under another ten ML-selected conditions showed actual discharge capacities converging to a narrower range between 223 and 229 mAh/g, thereby reducing the average predictive error to approximately 1.5%. While the highest predicted discharge capacity (229 mAh/g) was not experimentally attained, the tighter clustering around the high-capacity region underscores how iterative data integration systematically improves both accuracy and reliability.

Notably, these improvements were made possible largely by the deliberate and systematic incorporation of suboptimal or "failed" experimental outcomes into the iterative training process. Whereas traditional experimental methodologies tend to omit or inadequately document these suboptimal data points, thereby depriving ML models of crucial boundary information, the active learning approach leveraged such cases as essential feedback. Consequently, iterative active learning not only improved prediction accuracy but also provided a more comprehensive and balanced understanding of the parameter space and performance boundaries.

**Reducing Anthropogenic Bias and Enhancing Model Interpretability through Iterative Active Learning**

Beyond improving predictive performance, the iterative active learning framework also provides an advantage of expanding parameter distributions. **Figure 5** illustrates the distributions of key design parameters under four distinct scenarios: the original heuristic-driven dataset, a control scenario without active learning, and the first- and second-iteration active-learning proposals. Notably, ML-driven active learning frequently suggested parameter combinations that traditional domain expertise would typically overlook or dismiss—such as substantially different coating temperatures or uncommon co-dopant ratios.

While some of these ML-proposed conditions inevitably yielded suboptimal or even poor outcomes, such results served as valuable "negative data," significantly enriching the dataset for subsequent retraining. By explicitly integrating these negative examples, the model gained clearer boundary conditions, steering future predictions away from truly ineffective parameter regions. In contrast to conventional OFAT methods that inherently prioritize proven successes and frequently neglect negative outcomes, active learning systematically incorporates both successful and unsuccessful data points. This comprehensive and objective approach enables the model to achieve a deeper, more balanced, and robust understanding of the parameter space, ultimately reducing bias and improving the overall reliability of predictive modeling.

## Conclusion

In this work, we demonstrated how iterative ML integrated with active learning can effectively address two perennial challenges in battery material optimization: data scarcity and anthropogenic bias. By systematically venturing into parameter regimes that domain experts often neglect, the active learning approach not only improved predictive accuracy but also uncovered high-performance conditions outside traditional OFAT boundaries. Critically, the inclusion of failed or suboptimal experiments—rarely documented in conventional studies—emerged as a key driver of improved model reliability and interpretability.

Future directions include expanding the dataset to encompass additional variables and systematically logging negative/failed experiments across multiple labs, thereby enhancing the transferability of the model. Multi-objective optimization, such as balancing energy density, cycle life, cost, and sustainability, stands as another promising avenue. Finally, robust data-sharing protocols that capture meta-data for all experiments, successes and failures alike, will further refine ML-driven strategies. Collectively, these advances will help realize a self-correcting, data-centric paradigm for battery engineering, where human expertise and machine intelligence collaborate to push the boundaries of energy storage science.

# Methods

**Experimental Methods**

Nickel–cobalt–manganese (NCM) hydroxide precursors and lithium sources (LiOH or $Li_2CO_3$) were obtained from a industrial partner, ensuring consistent compositional purity and particle sizes around 10 μm, as verified by inductively coupled plasma (ICP) and scanning electron microscopy (SEM). All dopant and coating materials were commercially sourced with sub-micron particle sizes. Dry mixing employed a FM-Mixer operating at low speed for 1 min followed by high speed for 5 min to prevent thermal buildup. Calcination proceeded in a 65 m Roller Hearth Kiln (RHK), where the temperature profile, dwell times, and gas flow were held constant across all samples. Each batch of approximately 4 kg was placed in mullite saggars, ensuring uniform exposure to the thermal conditions. After calcination, the resulting powders were coarse- and fine-milled, washed at 15 °C with deionized water using a polypropylene filter press, and dried overnight under vacuum.

Coating procedures involved secondary mixing of the calcined powder and coating precursors in a FM-Mixer, followed by an additional heat treatment step in the same RHK but at a reduced load of 1 kg per saggar. The coated material was then coarse-milled and ultrasonically classified through a fixed mesh size.

Cathodes were prepared by mixing the active material (96.5 wt%), super C65 carbon black, and a polyvinylidene fluoride (PVdF) binder (12 wt% in NMP) to produce an electrode loading of ~10 mg·cm$^{-2}$. After casting, the electrodes were roll-pressed to a density of approximately 3.4 g·cm$^{-3}$ and punched into 14 mm disks. All electrochemical tests were conducted in CR2032 coin cells assembled with lithium metal foil (16 mm diameter) as the counter electrode, a 20 μm porous separator, and an electrolyte of 1 M $LiPF_6$ in EC/DMC (3:7 v/v). Cells were evaluated at 25 °C on a cycler using a constant-current protocol of 0.1C for both charge and discharge, defining 1C as 210 mAh·g⁻¹, and imposing voltage cutoffs of 4.25 V (charge) and 2.75 V (discharge). Residual lithium content, pH, tap density, and crystallographic properties from X-ray diffraction (XRD) were monitored to confirm process stability and identify potential anomalies, though such metrics were not employed directly in the machine-learning (ML) workflow.

**Data Preparation**

Experimental data spanning multiple years of NCM-based cathode syntheses were compiled into a single dataset that included design variables such as elemental composition, dopant amounts, calcination temperature, coating conditions, and process atmosphere. Observed electrochemical metrics—chiefly initial discharge capacity—were aggregated under standardized measurement protocols. Special care was taken to capture suboptimal and failed experiments, thereby mitigating selection bias. After discarding or imputing incomplete entries and validating metadata consistency, the resulting dataset was split into training and test subsets.

**Machine Learning and Active Learning Approaches**

A previously established pipeline guided the model selection and training procedures. Four algorithms—Decision Tree (DT), Random Forest (RF), Gradient Boosting Machine (GBM), and Neural Network (NN)—were compared using root mean square error (RMSE), mean square error (MSE), and $R^2$ as performance metrics. GBM showed superior predictive capability, particularly for skewed data with incomplete regions of parameter space. A fivefold cross-validation approach was used alongside a random-search hyperparameter routine. Hyperparameters varied over broad ranges, including subsample (0.01 to 1.0), n_estimators (50 to 300), max_depth (3 to 8), learning_rate (0.001 to 0.2), and min_samples_split (2 to 9).

**Domain Adaptation Framework using Particle Swarm Optimization**

An active learning (AL) loop for domain adaptation was devised to address data scarcity and anthropogenic biases. The GBM model first identified regions of high predictive uncertainty or sparse coverage; subsequent targeted experiments probed these parameter regimes, including doping concentrations and calcination conditions that domain experts traditionally excluded. Newly generated data, especially "failed" or boundary cases, were then fed back to retrain the GBM. Iterating through one-shot, two-shot, and higher AL cycles progressively improved predictive accuracy across a broader design space.

To refine the search for optimal parameter sets, a Particle Swarm Optimization (PSO) algorithm was employed, targeting maximal initial discharge capacity. Variables such as calcination temperature were incremented in 5 °C steps, coating time in 30 min steps, and dopant concentrations in 0.0005–0.001 increments (with $Nb_2O_5$ fixed at 0.002 wt% in selected experiments). Air atmosphere was maintained throughout. Over 1000 iterations, each exploring 100 candidate solutions, the algorithm selected the top 10 configurations predicted to yield the highest discharge capacity.

**Figure 1.** Overview of Conventional LIB battery fabrication (a) manufacturing chains, (b) cathode fabrication steps and its variables

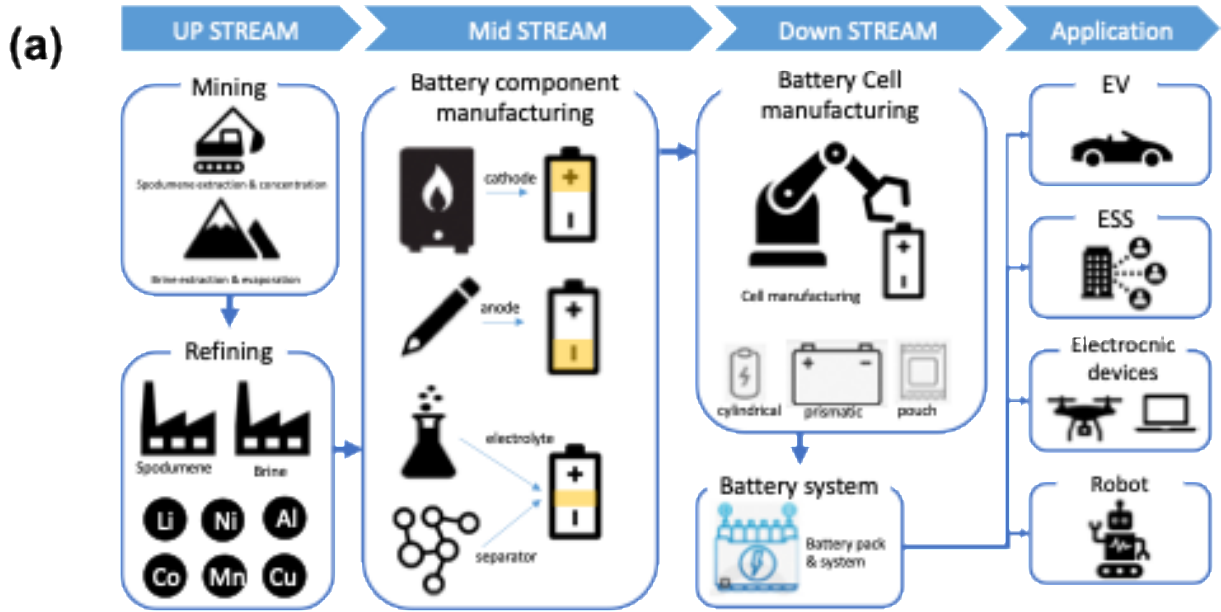

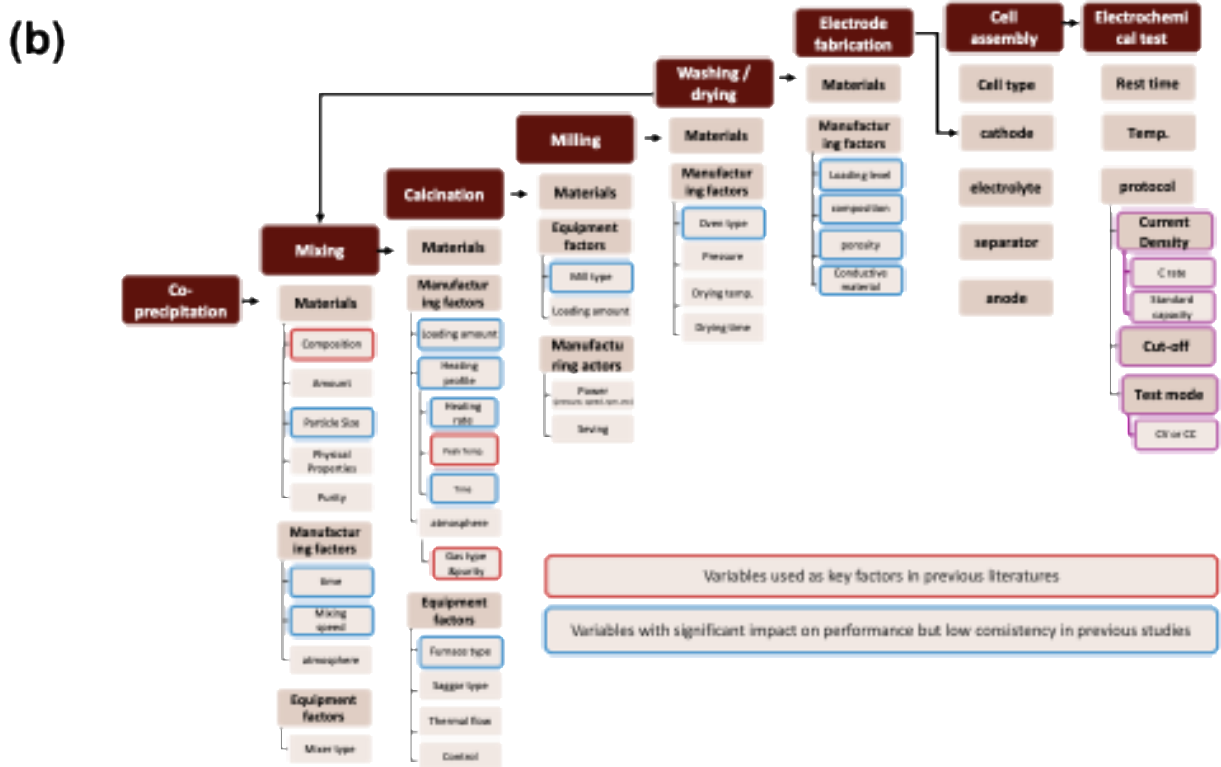

**Figure 2.** Distributions of key parameters: (left) near-Gaussian distribution for synthesis temperature; (right) highly skewed distribution for coating temperature.

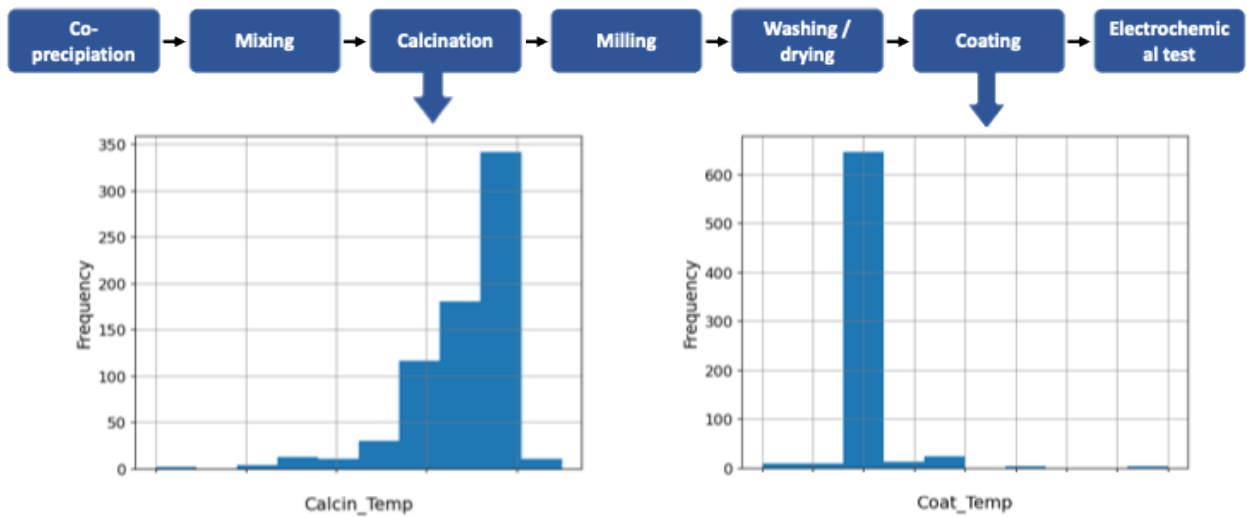

**Table 1.** Zero-shot performance of DT, RF, GBM, and NN in predicting initial discharge capacity.

|  | Dataset | DT | RF | GBM | NN |
|---|---|---|---|---|---|
| **RMSE (mAh/g)** | Train | 2.398 | 2.003 | 1.219 | 1.161 |
|  | Test | 2.398 | 2.207 | 1.65 | 1.601 |
| **R2** | Train | 0.833 | 0.884 | 0.957 | 0.961 |
|  | Test | 0.833 | 0.85 | 0.916 | 0.921 |
| **MSE** | Train | 5.750 | 4.012 | 1.486 | 1.348 |
|  | Test | 5.750 | 4.871 | 2.723 | 2.563 |

**Figure 3.** Active learning accuracy improvements: (a) zero-shot for the original dataset, (b) first active learning (one-shot)

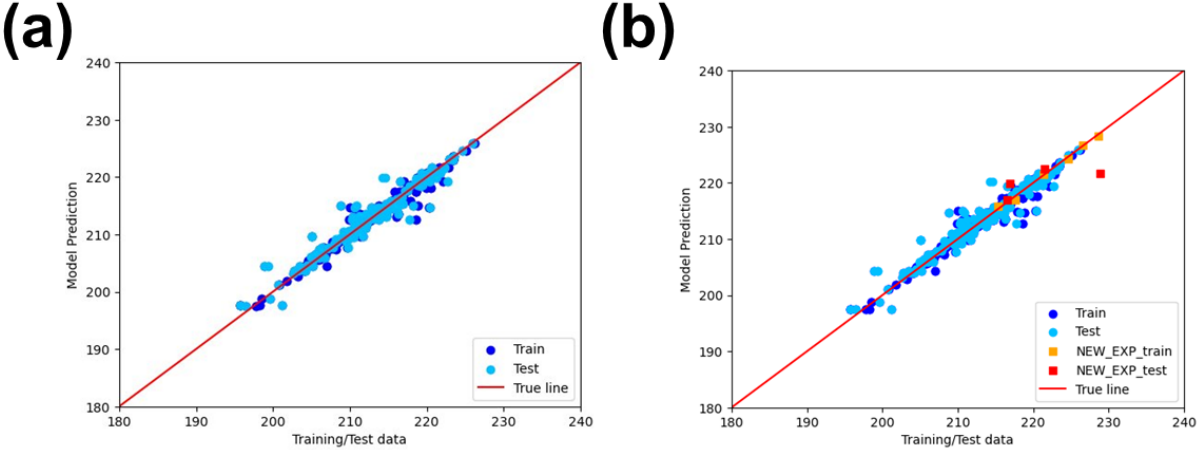

**Figure 4.** Correlation between model-predicted and experimentally measured discharge capacity over two active-learning cycles. Grey triangles depict the preliminary GBM model trained (△) and test (▽) data. Magenta open circles correspond to the ten "one-shot" experiments proposed by the first active-learning iteration, while blue filled stars represent the ten "two-shot" experiments proposed after retraining on the full one-shot results. The progressive clustering of points along the ideal 1:1 line (dashed) shows how iterative feedback reduces the mean absolute error from ~2.5 % to ~1.5 % and drives exploration toward a narrow, high-capacity window.

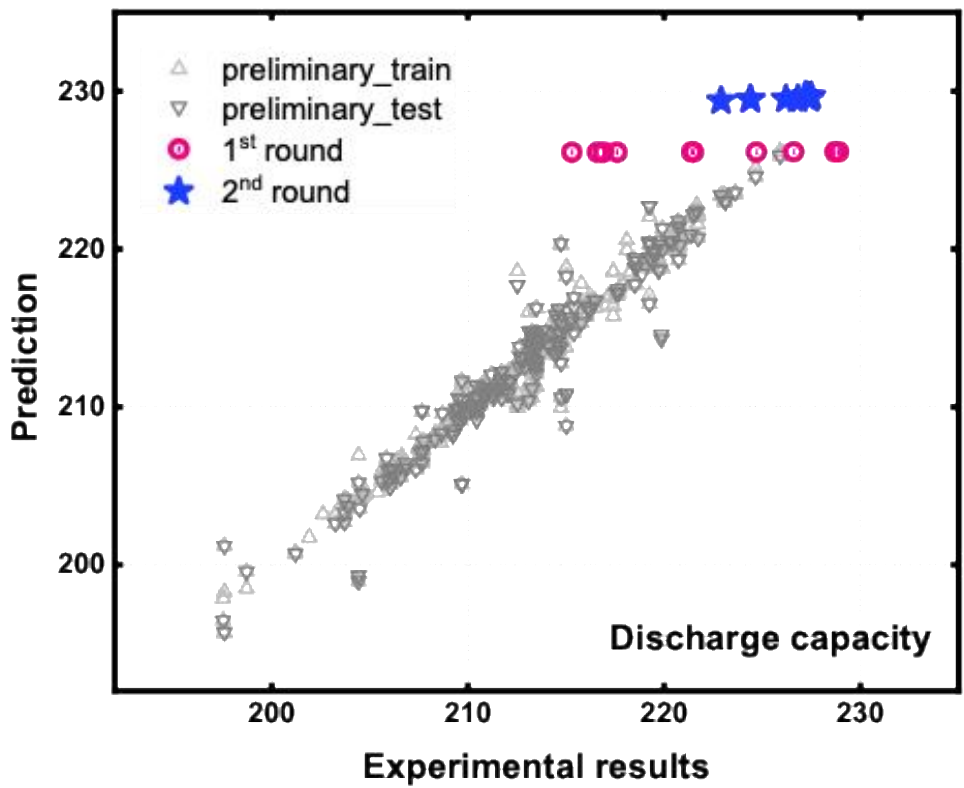

**Figure 5.** Comparison of selected parameters under ML- and domain expert-based experiment designs. The black points represent database values, while the red points indicate domain expert-based experimental results. The blue and green points correspond to the first and second active learning results, respectively. (a) Discharge capacity as a function of coating temperature, (b) discharge capacity as a function of calcination temperature.

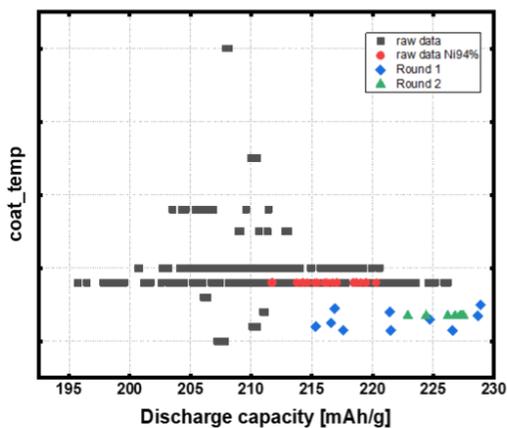 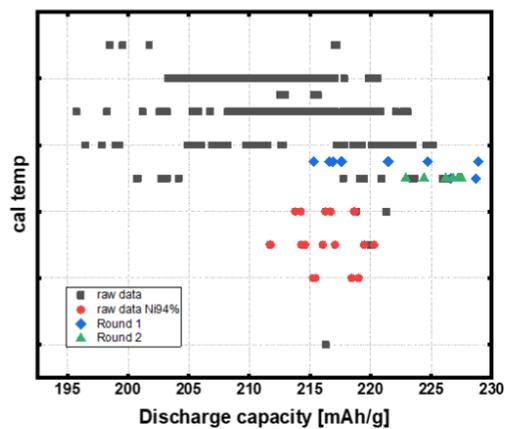